\documentstyle[epsf]{article}
\begin{document}
 
\parskip 2mm plus 1mm \parindent=0pt
\def\cl{\centerline}\def\lel{\leftline}\def\rl{\rightline}
 \def\hs1{\hskip1mm} \def\h10{\hskip10mm} \def\hx{\h10\hbox}
\def\vs{\vskip3mm} \def\vup{\vskip-2mm}
\def\page{\vfill\eject}
\def\<{\langle} \def\>{\rangle} \def\br{\bf\rm}  \def\it{\tenit}
 \def\de{\partial} \def\Tr{{\rm Tr}} \def\dag{^\dagger} 
\def\half{{\scriptstyle{1\over 2}}}
\def\ne{=\hskip-3.3mm /\hskip3.3mm}\def\tr{{\rm tr}}
\def\Pr{{\br Pr}}\def\to{\rightarrow}\def\N{{\cal N}}
\def\vb{\vskip20mm}\def\vm{\vskip10mm}
\def\cite{}

\def\be{\begin{equation}}\def\ee{\end{equation}}

\vskip10mm
\cl{\bf Speakable and unspeakable after John Bell} 
\vs\vs
\cl{\it A talk given at the International Erwin Schr\"odinger
Institute, Vienna (ESI)}
\vs\cl{at the November 2000 Conference in commemoration of John Bell}
\vs\vs\vs
\cl{2000 Dec 05}

\vskip10mm
\cl{Ian C. Percival}
\vskip7mm
\cl{(i.c.percival@qmw.ac.uk)}
\vskip15mm

{\bf Abstract} `Philosophy' was speakable for John Bell but is not for
many physicists.  The border between philosophy and physics is here
illustrated through Brownian motion and Bell experiments.
`Measurement', however, was unspeakable for Bell.  His insistence that
the physics of quantum measurement should not be confined to the
laboratory and that physics is concerned with the big world outside leads
us to examples from zoology, meteorology and cosmology.

\page

{\bf Philosophy and physics}

I first met John and Mary Bell at CERN because of our common
interest in the applications of nonlinear dynamics, but late
in life I became sufficiently fascinated by quantum philosophy
to follow John Bell's seminal contributions to that field, his 
`hobby'.
 
`Philosophy' is an unspeakable word for many physicists, but it was
not for John Bell.  He turned the question of local reality, or
nonlocality, from a philosophical question to one of physics, by
proposing an experimental test, the breaking of inequalities between
quantities that are measurable in the laboratory.

The boundary between philosophy and physics is not always clear.  Take
the earlier example of Brownian motion: `` Besides, there is a further
reason why you should give your mind to these particles that are seen
dancing in a sunbeam: their dancing is an actual indication of
underlying movements of matter that are hidden from our sight.  There
you will see many particles under the impact of invisible blows
changing their course and driven back upon their tracks, this way and
that, in all directions.  You must understand that they all derive
this restlessness from the atoms.  It originates from the atoms, which
move of themselves. \dots So the movement mounts up from the atoms and
gradually emerges to the level of our senses ...''
 
If you wanted to explain Brownian motion to a poet in the twenty-first
century, you couldn't do much better than this.  But it was not
written in this century.  It was written in Latin verse {\it by} the
poet and Epicurean philosopher Lucretius, before 55BC. [1]

Despite the clarity and accuracy of the description, the poem shows no
evidence of physics as we know it since the Renaissance.  For Brownian
motion, Brown's systematic studies were published in 1828, following
earlier observation using microscopes.  The comparison of
the detailed theory of Einstein and the experiments of Perrin
in the early years of the twentieth century finally established
for almost all physicists the reality of atoms.

Physics has many sides, theoretical and experimental, and one
important part of it is asking the right questions.  Ancient
philosophers like Lucretius did that, and modern
physicist-philosophers like Einstein, Bell and Shimony have done it
too.  They have survived on dangerous ground, and built a bridge
between unspeakable philosophy and speakable physics.

In physics we are required to put the answers to these questions to
the test, particularly experimental test, and here there is a parting
of the ways.  Lucretius could not distinguish the atomic hypothesis
from a continuum hypothesis, or any other.  Lucretius's was a
philosopher's poem.  There was keen observation, but to our knowledge
it was not systematic, and there was no attempt to vary the
conditions.  His explanation was plausible, but he could not check the
validity of the atomic hypothesis.  That did not happen for another
1950 years.

In providing his tests of nonlocality, Bell showed how
the implicitly philosophical considerations of Einstein, Podolsky and
Rosen [2] might be tested in the laboratory.  Like the theoretical
physicists who studied Brownian motion, he turned the philosophy into
physics.

\vs
{\bf Black boxes and nonlocality} 

Einstein [3] and Bell [4] were
particularly concerned with the question of nonlocal causality, in
which cause and effect are spatially separated in spacetime, so that a
signal from cause to effect would have to go faster than the velocity
of light.

According to classical special relativity, causality can act forwards
in time, but it does not act nonlocally over intervals that are
spatially separated.  An event can affect a future event, in or on its
forward light cone, but not a spatially separated event.  Consequently
signals cannot be sent faster than light.
  
But apparently, according to quantum theory, classical events that are
linked by quantum systems are different.  For them, there is a sense
in which causality might act nonlocally, but without any signalling
faster than light.  This is Bell's weak nonlocality, which can be
formulated in terms of the inputs and outputs of black boxes. 

John Bell in `Against Measurement' [5], discussed the possibility of
an exact formulation of some serious part of quantum mechanics: ``By
`serious' I mean that some substantial fraction of physics should be
covered.\dots I mean too, by `serious' that `apparatus' should not be
separated out from the rest of the world into black boxes, as if it
were not made of atoms and not ruled by quantum mechanics'
Nevertheless, it helps to analyse an experiment to test Bell
inequalities as a black box containing classical parts and a quantum
system.

An electrical engineer's black box consists of a circuit with input
and output terminals.  He may not know what circuit is inside, but it
is assumed here to be classical.  If there is no noise in the circuit,
then the black box is deterministic.  The outputs $j$ then depend on the
inputs $i$ through a unique transfer function $F$,  where
$$
j = F(i)
$$
and by experimenting
with different inputs and looking at the outputs, engineers can find
$F$.

In practice the resistors in the circuit produce noise, which we
assume to be classical noise.  The system is then stochastic.  The
noisy circuit can be represented by a probability distribution
$\Pr(F)$ over the transfer functions $F$, in which the unknown values
of supposedly classical background variables, like the coordinates of
thermal electrons, determine the particular $F$ that operates.

A physicist's black box contains an evolving physical system, such
as a classical electrical circuit, or an entangled quantum state with
classical inputs and outputs.  She may not know what physical system
is inside, but by experimenting with different inputs and looking at
the outputs, she can find out something about it.  

A Bell experiment is an example of a black box with classical
terminals and an entangled quantum system inside.  We suppose that the
source of entanglement is inside the black box, not an input.  For
photon polarization the setting of the orientations of the polarizers
is an input, and the detection of the directions of polarization is an
output.  All the inputs and outputs are classical events.

If we ignore backward causality, special relativity distinguishes
between two types of deterministic system, those with local transfer
functions $F$ for which the influence of an input on an output goes at
no more than the velocity of light, and those with nonlocal transfer
functions, for which the influences can act over spacelike
intervals.  It is possible to determine whether the transfer function
of a system is local or not by experimenting with different values of
the inputs, and observing the outputs.  There is no need to look
inside the black box.  All classical systems have local transfer
functions, as required by special relativity.


When classical or quantum systems are stochastic, and the inputs are
given, the probabilities of the outputs can be obtained from a
probability distribution $\Pr(F)$.  The stochastic systems are of
three main types, with different locality properties.

In the first type only local $F$ contribute.  It is therefore not
possible to send signals faster than the velocity of light.  
For the second type, which may soon be seen, transition probabilities
can only be obtained from $\Pr(F)$ in which at least one nonlocal
transfer function has nonzero probability, so there is an element of
nonlocality.  But nevertheless it is not possible to send signals
faster than the velocity of light.  The system is then weakly
nonlocal, or nonlocal in the sense of Bell.  The {\it definition}
of weak nonlocality needs no quantum theory.
For the third type, which we never expect to see, it is possible to send
signals faster than the velocity of light.

The stochasticity of classical systems comes from background variables
that are not included in the system, but for quantum systems it does
not come from any background variables that we can see, so either they
are assumed not to exist, as in the Copenhagen interpretation, or they
are called hidden variables.

In a Bell experiment in which the entangled quantum system is
sufficiently close to a pure state, and the measurements sufficiently
good, the black box is weakly nonlocal in the sense of Bell.  An
experimenter who has never seen the apparatus before can then tell by
experimenting with the inputs and outputs, and without looking inside,
that the black box contains a quantum system.  This property of black
boxes containing quantum systems comes from weak nonlocality.


So these black boxes tell us something about the world: there are
correlations between classical events that can only be produced by
quantum links.  These correlations are important in their own right.
They are weakly nonlocal.  They also show that the properties of our
world cannot be explained using local hidden variables, but that is
not their main significance.

So weak nonlocality is important for all physicists, whether they are
interested in hidden variable theories or not.  Weak nonlocality is
unique in modern physics: classical dynamics, quantum dynamics and
classical general relativity are all local.  Nonlocality only occurs
in some of those processes for which quantum states influence
classical events.  Laboratory quantum measurement is such a process,
but it is not the only one.

Today only some ideal experiments involving quantum measurement are
nonlocal in any sense, though there may soon be real experiments.

\vs
{\bf The nature of Bell experiments}

There is a profound distinction between experiments to test weak
nonlocality by the violation of Bell inequalities and most other
experiments with quantum systems.  

Consider for example an experiment to determine the spectrum of an
atom, or a differential cross section for the scattering of an
electron by a molecule, or an experiment to determine the band gaps of
a solid, or to find a new particle.  The aim of all these experiments
is to determine the properties of quantum systems.  The classical
apparatus used to prepare the system and to make the necessary
measurements is essential, but secondary to obtaining these properties.

In Bell experiments the converse is true.  The aim is to test for
violation of the inequalities, which are derived from the
(statistical) properties of classical events, such as the setting of
the apparatus, which is a classical input, or the detection of a
particle by an electron avalanche, which is a classical output.  The
probabilities of the outputs, given the inputs, are what appear in the
Bell inequalities, and it is the location of these events in spacetime
that determine the locality or nonlocality.  

These classical events are connected by an ancillary quantum system,
whose function is to produce their unusual statistical properties.
The quantum properties, like the entanglement of particles, or the
polarization of photons, or the spins of atoms, are essential, but
secondary.  The primary result is the violation of an inequality.
The apparatus as well as the quantum system is essentially involved.



This distinction has implications for the analysis of Bell
experiments.  Real Bell experiments are designed to approximate ideal
experiments.  But the classical events in a real experiment are
usually different from those in the ideal experiment which it simulates.
There are generally more types of possible output.
For example, in an ideal experiment, it is usually assumed that the
detectors detect every particle, but in real experiments they
don't.
  
It follows that the inequalities of the ideal experiment do not
always apply directly to the real experiment, and further assumptions
are needed to demonstrate weak nonlocality.  

There are two ways to tackle this problem.

One is to perform those special experiments for which the ideal and
the real are the same.  In these the detectors do not distinguish
between some of the particle quantum states and a failure to detect a
particle.
Experiments to test the Clauser-Horne-Shimony-Holt inequality 
is of this type, so this inequality has been widely used.

Another way is to recognize that every real experiment that simulates
an ideal Bell experiment has its own critical inequalities, that apply
directly to the probabilities for all the outputs of the real
experiment.  
These can be derived from the condition that all the
$\Pr(F)$ are local.  The larger the number of detectors, the larger
the number of inequalities, and a computer program may be needed to
obtain them.


If any one of these inequalities is violated, weak nonlocality has
been demonstrated, and no further assumptions are needed.

\vs
{\bf Measurement}

was an unspeakable word for John Bell:

John Bell: Against `measurement' {\it Physics World} 33-40 Aug 1990,
p34:
`` When I say that the word `measurement' is worse than the others
\dots I do have in mind its use in the fundamental interpretive rules
of quantum mechanics. \dots
`` The first charge against `measurement', in the fundamental axioms
of quantum mechanics, is that it anchors there the shifty split
of the world into `system' and `apparatus'. ''
I will not be discussing this charge.  It does not apply to explicit
dynamical models of physical processes of which quantum measurement is
an example, and these are discussed by Gian-Carlo Ghirardi at this
meeting.

`` A second charge is that the word comes loaded with with meaning
from everyday life, meaning which is entirely inappropriate 
in the quantum context.  When it is said that something is
`measured' it is difficult not to think of the result as referring
to some preexisting property of the object in question.
This is to disregard Bohr's insistence that in quantum phenomena
the apparatus as well as the system is essentially involved. ''

This charge was avoided above by treating the quantum system and
classical apparatus together as a single system.

`` In the beginning natural philosophers tried to understand the
world around them.\dots Experimental science was born.  But
experiment is a tool.  The aim remains: to understand the world.
To restrict quantum mechanics to be exclusively about piddling
laboratory operations is to betray the great enterprise.  A serious
formulation [of quantum mechanics] will not exclude the big
world outside the laboratory. ''
 
I will discuss this third charge now.

\vs
{\bf The big world}

Traditionally quantum measurements take place in the laboratory, but
the laboratory is only part of our universe, and all such measurements
start out as imitations of natural phenomena.  

Cloud chambers were based on the physics of clouds, which are natural
detectors of charged particles.  Spark chambers imitate lightning.  We
can generalize {\it quantum measurement} to mean any process whereby
the state of a quantum system influences the value of a classical
variable.  This definition then applies to the big world.


Laboratory quantum measurements include particle states producing the
droplets in cloud chambers, bubbles in bubble chambers and sparks in
spark chambers.
They include photon states producing silver grains in photographic
emulsions, and also electron avalanches in solid state detectors and
photomultipliers.

Other quantum measurements include photon states sending
impulses through the optic nerves of owls, the states of cosmic rays
that produced small but very long-lived dislocations in mineral
crystals in the Jurassic era, and the quantum fluctuations that are
believed to have caused today's anisotropies in the universal
background radiation and in galactic clusters. 

This takeover of the physics of laboratory quantum measurement into
the world outside the laboratory is here generalized, and one of the
questions we have to ask is how far this generalization can go.
 
\vs
{\bf  Equilibrium gases}

Laboratory systems used for quantum measurement are very complicated
physical systems, even stripped down to their bare essentials.
They involve amplification in one form or another, and so do the
natural systems that they imitate.

A gas in equilibrium is much simpler, yet generalized quantum
measurement takes place there also.
The reason is that the motion of the molecules in the gas is chaotic,
and small changes now result in large changes later.  In particular
changes at the quantum level now produce significant classical
fluctuations in the density later.  However, unlike earlier examples,
we can't use the classical density fluctuations to learn anything
specific about these earlier quantum states, because the chaos causes
mixing, which effectively obscures the signal.

In the nineteenth century, Rayleigh recognized that these
classical density fluctuations would scatter light, and
that the scattering was strongly dependent on the wavelength of
the light.  The result is the blue of the sky.

The growth of droplets of water around the charged particles
produced by cosmic rays in the atmosphere is a quantum measurement.
So are the density fluctuations in the atmosphere that cause
the sky to be blue where there are no clouds.
So if you ever look at the sky, as every physicist sometimes should,
whether it is clear or overcast, you are seeing one example or another
of quantum measurement.

\vs
{\bf Theoretical cosmology makes a prediction}








According to the cosmologists there occurs in the early universe a
generalized quantum measurement.  Quantum fluctuations produce
fluctuations in classical variables and these are then rapidly
stretched from subatomic scales to the size of galaxies or even
larger. So according to this theory, the inhomogeneities in the
universe, such as galactic clusters, galaxies, and the fluctuations in
the background radiation are all due to quantum measurement in the
early universe.  Somewhat later, physicists meeting in Vienna were
also produced by fluctuations from the same process of quantum
measurement.

Recently there have been several detailed observations of the
inhomogeneities\hs1 in the universal background radiation,\hs1 for example
those of the `Boomerang' project [6].  The intensity of the fluctuations
were plotted as a function of the order of the spherical harmonics,
as illustrated in the figure.  There is a pronounced peak.

\begin{figure}[htb]
\epsfxsize14.0cm
\centerline{\epsfbox{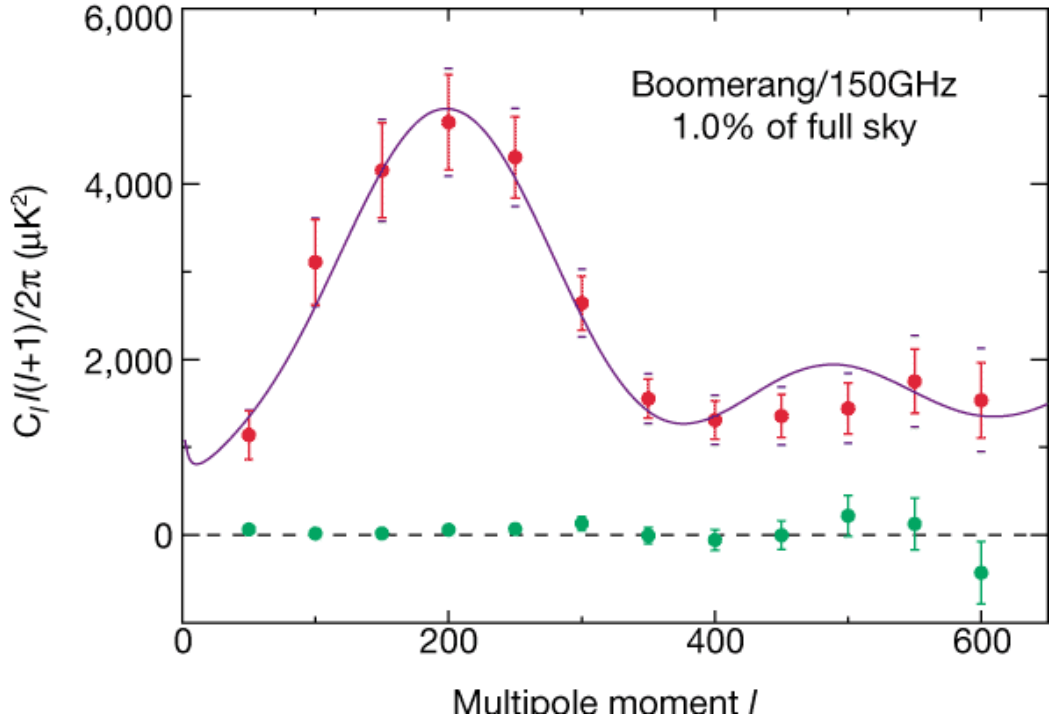} }
\label{fig}

\caption{Theory (continuous line) and experiment for inhomogeneities
in the universal background radiation}
\end{figure}

The observations are compared with theory, represented by the
continuous curve, and based on the assumption of a flat universe.  
The theory shows the same peak, and a number of smaller peaks
which are not (yet) seen.  By the standards of cosmology, the
agreement between theory and observation is excellent, providing
evidence for quantum measurement in the early universe.

\vs
{\bf Conclusions}

One of John Bell's major objections to quantum `measurement' might be
overcome by generalizing the definition to include processes in the
big world.  With this extended definition, quantum measurement is a
universal property of physical systems.  It is all around us, and we
would not be here without it.

Consequently the dynamics of quantum measurement has universal
significance.  So have its properties, like the weak nonlocality
of John Bell.

\vs
{\bf Acknowledgements} My thanks go to the ESI for their hospitality and
support, and to Mary Bell, Reinhold Bertlmann, Abner Shimony and Dave
Sutherland for amendments and corrections.

\vs\vs
{\bf References}

[1] Lucretius, Nature of the Universe (Trans. R.Latham, Penguin, Harmondsworth 1951) p53-54.

[2] A. Einstein, B. Podolsky and N. Rosen.  Can quantum-mechanical
description of reality be considered complete? {\it Phys. Rev.} {\bf
47}, (1935) 777-780.

[3] A. Einstein, Albert Einstein, Philosopher Scientist
(ed. Schilpp, Cambridge University Press, Cambridge, UK 1959) p85.

[4] J. S. Bell.  On the Einstein Podolsky Rosen paradox {\it Physics}
{\bf 1} (1964) 195-200.

[5] J. S. Bell.  Against `Measurement' {\it Physics World}, August
(1990) 33-35.

[6] De Bernadis et al. A flat universe from high-resolution maps of
the cosmic microwave background radiation. {\it Nature}
{\bf 404} (2000) 955-959.

\end{document}